
\input harvmac

\Title {DAMTP93/R14}
{{\vbox {\centerline{Anomalous Fermion Production in Gravitational Collapse
 }}}}

\bigskip
\centerline{G.W. Gibbons}
\centerline{Alan R. Steif}
\bigskip\centerline{\it D.A.M.T.P.}
\centerline{\it Silver St. }
\centerline
{\it Cambridge University}
\centerline{\it Cambridge, U.K. CB3 9EW }
\centerline{\it gwg1@moria.amtp.cam.ac.uk}
\centerline{\it ars1001@moria.amtp.cam.ac.uk}
\vskip .2in

\noindent
ABSTRACT: The Dirac equation   is solved in the Einstein-Yang-Mills background
found by  Bartnik and McKinnon.  We find a normalizable zero-energy
fermion mode in the $s$-wave sector. As shown recently, their solution
corresponds to a gravitational sphaleron which mediates transitions between
topologically distinct vacua. Since the Bartnik-McKinnon solution is unstable,
it will either collapse to form a black hole or radiate away its energy. In
either case,  as the Chern-Simons number of the configuration changes, there
will be an accompanying anomalous change in fermion number.

\Date{May 21, 1993}
\def\a{\alpha}
\def\b{\beta}
\def\ra{\rightarrow}
\def\o{\omega}
\def\m{\mu}
\def\p{\partial}
\def\th{\theta}
\def\t{\tau}
\def\s{\sigma}

\def\n{\nu}
\def\g{\gamma}
\def\pr{\prime}
\def\d{\bf d}
\def\bmk{Bartnik-McKinnon}
\def\cs{Chern-Simons}
\def\snt{{\rm sin}\, \theta}
\def\cst{{\rm cos}\, \theta}

\def\bo{{\bf \omega}}
\def\DT{\hbox{$ D_T\kern -12.5pt / \kern +12.5pt$}}
\def\sdn{\vec\s\cdot \vec n}
\def\tdn{\vec \t\cdot \vec n}
\def\ndt{ \vec n\cdot \vec \t}
\def\nst{  \vec n \cdot \vec \s \times \vec \t }
\def\P{\Psi}
\def\O{\Omega}

{\newsec{Introduction}}

Neither  the  Yang-Mills nor  Einstein field equations   admit static
finite-energy, non-singular solutions. However,
   such  particle-like solutions have been found by Bartnik and McKinnon
\ref\bm{
R. Bartnik and J. McKinnon, {\it Phys. Rev. Lett.} {\bf 61} (1988) 141.
}
in  the combined
Einstein-Yang-Mills theory.  These solutions were later interpreted as
sphalerons, that is, static saddlepoint solutions lying at the top of an energy
barrier in field configuration space separating vacua with different
Chern-Simons number
 \ref\gv{
D. Gal'tsov and M. Volkov, {\it Phys. Lett. } {\bf B273} (1991) 255.}
\ref\sw{D. Sudarsky and R. Wald, {\it Phys. Rev. D} {\bf 46} (1992) 1453.}.
 This, in particular,  accounts for their instability. If the sphaleron is
perturbed,   it will  either radiate its energy to infinity or collapse to form
a black hole. Since either process involves a change in Chern-Simons number,
one    expects an equal anomalous change in chiral fermion number.
In this paper we initiate a study of this problem. In Section
2, the {\bmk} solution and its interpretation as a sphaleron is reviewed.   In
Sections 3-5, the Dirac equation in the background of the static {\bmk}
sphaleron is analyzed.  We find  a zero energy bound state in the $s$-wave
sector.   In Section 6, a Higgs field is included. In Section 7, the conformal
invariance of the Dirac equation is exploited   to prove a general no-hair
theorem for fermions.

{\newsec{Bartnik-McKinnon Solutions}}

In {\bm}
 a class of spherically symmetric particle-like solutions to
the Einstein-Yang-Mills equations with  $SU(2)$ gauge group were found
numerically. It was shown that there are an  infinite sequence of solutions
$(A_n (r), B_n (r), K_n (r))$ for the metric and gauge field
\eqn\metric{
ds^2 =  -A^2 dt^2  + B^2 dr^2 + r^2 (d\th^2 + {\sin}^2\,\th d\phi^2)
}
 \eqn\wuyang{
A_i = { 1- K \over 2gr} \epsilon_{ijk} n^j \t^k
\quad \footnote{*}{\rm This form of the gauge field is gauge equivalent to that
used
in {\bm}. }
}
where $g$ is the gauge coupling constant, $\t^i$ are the generators of $SU(2)$,
and
 \break
$\vec n = (\sin\,\th\, \cos\,\phi, \sin\,\th\,\sin\,\phi, \cos\,\th).$
  $K_n$ has  $n$ nodes and  satisfies the boundary condition $K_n (0) =1 $ and
$K_n(\infty ) = (-1)^n $. At large distances, the metric approaches
Schwarschild and  the gauge  field
strength decays as $1/r^3$  with  $A_i$ approaching the pure gauge
$ A_i= - {i\over g} \p_{i} U U^{-1}$ where $U=1$ for $n$ even and  $U = -i \ndt
$ for $n$ odd.
  Near a node $r=r_0$ ($K (r_0) =0$),  the gauge field
corresponds to   a $ P = 1/g$ Dirac monopole, and  the metric to
 extremal Reissner-Nordstrom.
The masses of the {\bmk} solutions are proportional to   the only mass scale in
the problem,
$g^{-1} G^{-1/2}$, where $G$ is the gravitational constant.
 The masses increase with $n$ approaching the mass of the extremal
Reissner-Nordstrom black hole with magnetic charge, $P= 1/g $ as $n\ra \infty$.
Heuristically, in this limit, $K_n$ fluctuates rapidly approaching its mean
zero value corresponding to the Dirac monopole.
The existence of the {\bmk} solutions was subsequently established
rigourously
\ref\swy{
J. Smoller, A. Wasserman, and S. T. Yau, {\it Comm. Math. Phys.}
{\bf 143} (1991) 115.
},
 and generalized to include horizons
\ref\bizon{
P. Bizon, {\it Phys. Rev. Lett. } {\bf 64} (1990) 2844.
}.

 Soon after their discovery, it was shown that both  particle-like and black
hole solutions are unstable
{\ref\sz{
N. Straumann and Z. Zhou, {\it Phys. Lett. } {\bf B237} (1990) 353;
N. Straumann and Z. Zhou, {\it Phys. Lett. } {\bf B243} (1990) 33.
}}.
In fact, the Bartnik-McKinnon solutions
 correspond to a gravitational analog of  the sphaleron {\gv  \sw}.
We recall that sphalerons {\ref\nm {N. Manton,  {\it Phys. Rev. D} {\bf 28}
(1983) 2019;
 F. Klinkhamer and N. Manton,  {\it Phys. Rev. D} {\bf 30} (1984) 2212.}
  are static saddle-point  solutions
that lie at the top of an energy barrier
separating topologically distinct vacua.
For asymptotically flat spacetimes, the ADM energy provides a positive definite
mass functional on field configuration space. The zero energy ``vacua" are flat
spacetime metrics and pure gauge  Yang-Mills configurations
$ A_i= - {i\over g} \p_{i} U U^{-1}$. Each pure gauge yields a map from space
into the group manifold of $SU(2).$
Demanding $U\ra 1$, this becomes a map $U: S^3\ra S^3 .$  Since maps with
different winding,  or Chern-Simons, numbers cannot be continuously deformed
into one another, paths in configuration space connecting these vacua must
enter non-vacuum regions, and therefore by the positive definiteness of the
mass functional must traverse an energy barrier.
 \footnote{*}{\rm We shall regard   vacua which are related by so called large
gauge transformations as distinct. One can, if one wishes, identify them, in
which case the paths we refer to below become closed non-contractible loops in
this identified configuration space.}  Each such path has a maximal energy
configuration.  Among these configurations the one  with minimum energy   will
be  an extremum of the mass functional  and therefore correspond to an unstable
saddlepoint solution.   By considering paths connecting the  vacuum sector with
Chern-Simons number zero and the sector with Chern-Simons number one, one
obtains  the { \bmk } solution  $K_1.$ One can see
by symmetry that  its Chern-Simons number is $1/2$. Repeating this procedure,
with paths connecting $K_1$ and its gauge equivalent solution with {\cs} number
$-1/2$,   the $K_2$ solution with zero {\cs} number is obtained. Continuing in
this
way, the entire {\bmk} sequence is produced. This procedure also yields the
black hole solutions if one considers asymptotically flat  metrics with  fixed
horizon area.

As the fields evolve dynamically along a path in configuration space connecting
vacua with different Chern-Simons numbers and  passing  through the sphaleron,
anomalous fermion production will occur {\ref\thooft{G. 't Hooft,  {\it Phys.
Rev. Lett.} {\bf 37} (1976) 8;
 {\it Phys. Rev. D} {\bf 14} (1976) 3442.}.} The total number of chiral
fermions created equals the change in the {\cs} number of the field
configuration. As one passes between topologically distinct vacua, a fermion
state will emerge from the negative energy Dirac sea, enter the discrete
spectrum, and finally  merge with the positive energy continuum.  By symmetry,
therefore, there should be a zero energy bound state at the midpoint of this
path where
the sphaleron configuration is located.
If perturbed,  the sphaleron    will  fall in one of two directions. The
Yang-Mills field will dominate, in which case the sphaleron will radiate away
its energy to  infinity. The  fermion zero mode would then also escape to
infinity perhaps after being reflected at the origin.  Alternatively, gravity
may induce the object to collapse to a black hole swallowing the Yang-Mills
field, and presumably the fermion mode as well.

{\newsec{Dirac Equation}}

We now proceed to test this picture by solving the Dirac equation
in the sphaleron background.  As expected from the above discussion,
we find a zero-energy bound state in the $s$-wave
sector. We also derive the full set of time dependent radial equations.    To
determine more precisely the fate of the zero mode if the sphaleron collapses
to a black hole would require numerically solving the time dependent equations.
Such an analysis was done for the  Yang-Mills-Higgs sphaleron in
{\ref\bri{J. Kunz and Y. Briyahe, {\it Phys. Lett. } {\bf B304} (1993)
141.}}{\ref\cott{ W. Cottingham and N. Hasan, {\it Nucl. Phys. B} {\bf 392}
(1993) 39.}}.

The massless Dirac equation
in an Einstein-Yang-Mills  background for an isodoublet fermion is given by
\eqn\diraceq{
i\g^{\m} (\nabla_{\m} -i g A_{\m}) \P   =0
}
where  $ \nabla_{\m}$ is the covariant derivative $\nabla_{\m}  =
 \p_{\m} - {1\over 4} \o_{\m}^{ab} \g_a
\g_b .$ $\m$ and $a$ are  tangent  and spacetime  indices
respectively and are related by $e^a_{\m} \equiv {\bf e}^a, $ a basis of
orthonormal  one-forms.
$\o_{\m}^{ab} \equiv {\bf \o}^{ab}$ are the associated  connection
one-forms  obeying  ${\bf d  e}^a +
{\bf \omega }^a_{\,\, b} \wedge {\bf e}^b =0 ,$ and
$\g^a$ are Dirac
matrices satisfying $\{ \g^a, \g^b \} = -2\eta^{ab} $  with $\eta^{00} = -1$.
  The metric and gauge field are given in {\metric} and {\wuyang} where $\t^i$
are now the Pauli spin matrices. In terms of the orthonormal basis
\eqn\basis{
{\bf e}^0 = A {\bf d}t \quad
{\bf e}^r = B {\bf d}r \quad
{\bf e}^{\th} = r {\bf d}{\th} \quad
{\bf e}^{\phi} = r {\rm sin}\,\th{\bf d}\phi
}
 the connection one-forms are
\eqn\connection{
\bo^{0r} = { B^{-1} A^{\pr} } {\d}t\quad
{\bo}^{\theta r} = {B^{-1}} \d \th\quad
{{\bo}^{\phi r}} = {B^{-1}}{\snt}  \d \phi\quad
\bo^{\phi \th} = \cst \d \phi
}
where $\pr \equiv  {\p\,\over \p r}$.
In  the chiral representation, the gamma matrices are
\eqn\gammas{
\g^0 = {\pmatrix{0&1\cr 1 & 0\cr}}\quad\quad
\g^i =  {\pmatrix{ 0 & \s^i \cr -\s^i & 0\cr}}\quad\quad
\g^5 = {\pmatrix{ -1 &0\cr \,0& 1\cr}}
}
where $\s^i$ are Pauli spin matrices.  $\P$ can be decomposed into its left and
right chiral components $\P = {\pmatrix { \psi_L\cr \psi_R\cr}}$
where $\psi_{L(R)}$ carry two-component Lorentz and isospin indices.
Since the Dirac equation is massless, the two chiralities decouple. In the
following, we consider just the right-handed component $\psi\equiv \psi_R$. The
Dirac equation then becomes
\eqn\diracb{
  {\p \psi  \over \p t} +
 {\sdn \over r}{ A^{1/2} \over B }
 {\p\,\over \p r} (rA^{1/2} \psi)  +
{A \over r} \DT
\psi
 -  ig A \vec \s \cdot \vec A \psi    = 0
}
where $\DT$ is the Dirac operator on the unit two sphere and  $ \vec a\cdot
\vec b \equiv a^i b^i$. Eqn. {\wuyang } implies
 $\vec \s \cdot \vec A= {(K-1)\over 2 g r}( \nst ) .$
 The conserved inner product is given by
\eqn\innerproduct{
<\psi_1| \psi_2> =  \int {\psi_{1}}^{\dagger }\psi_{2}  \; Br^2 dr d\Omega .
}

We now discuss the effect of charge conjugation, $C$, and
parity, $P$ on the sphaleron.  The action for
a Dirac fermion coupled to a Yang-Mills field
with arbitrary gauge group is invariant under $C$ and $P$ separately.
Under $C$, the fields transform as
\eqn\ctrans{\eqalign{
\P & \ra \P^C = \g^2 \P^*\cr
A_{\m} & \ra A_{\m}^C = - A_{\m}^*\cr
}}
and under $P$ as
\eqn\ptrans{\eqalign{
\P(x^i,t)  & \ra \P^P =  \g^0 \P (-x^i,t)\cr
A_{i} (x^i,t) & \ra A_i^P = - A_{i} (-x^i,t) \cr
A_0 (x^i, t) & \ra A_0^P = A_0 (-x^i,t)  .\cr
} }
The metric being real and spherically symmetric is invariant under $C$ and $P$.
Because $C$ and $P$ interchange  chirality,
the action for
chiral fermions  is only invariant under the combined
$CP$ transformation.
Using {\ctrans ,} {\ptrans}, and  $-\t^{i*} = \t^2 \t^i \t^2 ,$   the gauge
field {\wuyang} transforms
as $A_i \ra  A_i^{CP} = \t^2 A_i \t^2 .$ The $CP$ invariance of the theory then
implies that given a solution  $\P$ to the Dirac equation
in the sphaleron background  with  energy $E ,$  $\t^2 \P^{CP}$ is
a solution  in the same background, but with energy $-E$.

{\newsec{$S$-wave Sector and Zero Mode}}
  We now solve the  Dirac equation {\diracb} by separating variables.
Since the total angular momentum
$\vec K = \vec L + \vec S + \vec T$ commutes with the Hamiltonian,   $\psi$ can
be expanded in eigenstates of
 $K^2$ and $K^z$
with eigenvalues $k$ and $m. $ $\vec L$,    $\vec S = \vec \s /2$,  and $ \vec
T = \vec \t/2$ are the orbital angular momentum, spin, and isospin.
The $s$-wave sector corresponding to $k=m=0$
is spanned by the
two states $\chi_1$ and $ \chi_2 \equiv \sdn \chi_1$
where
\eqn\hedgehog{
 \chi_1 = { 1\over \sqrt 2} \biggl\lbrack
\pmatrix{1\cr 0\cr}_S \pmatrix{0\cr 1\cr}_T -
\pmatrix{0\cr 1\cr}_S \pmatrix{1\cr 0\cr}_T \biggr\rbrack
}
  is the hedgehog spinor
 satisfying
$(\vec \s + \vec \t) \chi_1 = 0 .$
The action of the various   operators in the Dirac equation on the  two states
can be determined from the hedgehog property and the spin commmutation
relations.
The transverse Dirac operator  can be written as $ \DT = (2 \vec S\cdot \vec L
 +1 )(\sdn) = (J^2 - L^2 - S^2 +1 ) (\sdn )$
where  $\vec J \equiv \vec L + \vec S$. Since
$[J^2 , \sdn] =0$ and    $\sdn$ changes the value of the orbital angular
momentum by one, we find that $ \DT   \chi_1 = -   \chi_2$ and   $\DT    \chi_2
  = \chi_1 .$
The operator $\nst$ appearing in the gauge field term in the Dirac equation
also interchanges $\chi_1$ and $\chi_2$:
 $(\nst )\chi_1 =  -2i  \chi_2$ and $ (\nst ) \chi_2 = 2i \chi_1 . $  Thus,
{\diracb} becomes
\eqn\swaveequations{\eqalign{
{\p f \over \p t} &      +  {\p g \over \p r*} +
 {AK\over r} g   = 0\cr
{\p g \over \p t} &  +  {\p f \over \p r*} - {AK\over r}      f
  = 0 \cr
}}
where $\psi = r^{-1} A^{-1/2} f \chi_1 +  r^{-1} A^{-1/2} g  \chi_2 $
and  $r^*$  is the ``tortoise" coordinate    satisfying  ${dr*\over dr} =
{A^{-1} B} . $

Eqn. {\swaveequations} admits a zero energy bound state of the form
\eqn\zeromode{
f =  \exp{  \int_{r_0}^{r} B {K \over r}  \, dr} \; ,\quad g =0
 }
 for $K = K_n$ with $n$ odd.  For $n$ even, $K \ra 1$ asymptotically, and
therefore $f$ diverges.  The zero mode {\zeromode} vanishes at $r=0$ and falls
off as $1/r$ asymptotically. The charge density  of the wave function from
{\innerproduct} is given by $ 4 \pi A^{-1}  B|f|^2 .$ The density  of the zero
mode {\zeromode} is peaked in the monopole region, $K=0$, which is effectively
acting as a potential well.
 For a spacetime with    horizon at $r=r_H$, the metric functions behave as
 $ A \sim (r-r_H)^{1/2} (1+ O(r -r_H))$ and
$B \sim (r-r_H) ^{-1/2}(1+ O(r - r_H))$. From {\zeromode} we observe that the
wavefunction diverges as
  $ (r-r_H) ^{-1/4}$, and  its norm   logarithmically. This is perhaps to be
anticipated in light of the no-hair theorem
for fermions.
 The effective potential
within the $s$-wave sector is found by rewriting
  the two coupled first order equations
as decoupled second order equations
\eqn\secondordereqn{
{\p^2 f \over \p t^2} -{\p^2 f \over \p r^{*2}}
+  V(r) f =0  ,\quad \dot  g  = - {\p f \over \p r*}  + H f}
where
\eqn\effpotl{
V(r) = {\p H\over \p r^*} + H^2, \quad H \equiv A{K\over r}
.}
Near the horizon, the effective potential
vanishes as $V \sim (r-r_H)^{1/2} .$

{\newsec{$k>0$  Modes}}
The eigenspaces with eigenvalues $k$ and $m$ are
four-dimensional for    $k \geq 1.$  Since $L^2$    and $K^i$  commute, they
can be simultaneously diagonalized. Within the four dimensional eigenspace,
there are   two states denoted
$|k,m, \pm >$ with  $L^2$ eigenvalues $ l= k \pm 1 $, and an  orthogonal
two-dimensional
space with  eigenvalue $ l =k$.  In addition, either  $J^2 = (L+S)^2$ or
$R^2\equiv (S+T)^2$ can be diagonalized, but not simultaneously since they do
not commute. The eigenstates of $J^2$ are  $|k,m, \pm >$
 with eigenvalues $j = k \pm 1/2$ and
 $|k, m, j_{\pm}>$, lying in the $l=k$ subspace, with  eigenvalues $j = k\pm
1/2. $ The eigenstates of $R^2$ are   $|k,m, \pm >$
 both with eigenvalues $r=1$ and
 $|k, m, 0>$ and $|k, m, 1>,$ lying in the $l=k$ subspace, with  eigenvalues $
r= 0$ and $1$.
The $J^2$ and $R^2$ eigenstates in the $l=k$ subspace are related by a rotation
matrix
of angle $\xi$ with $ {\tan}\, \xi = \sqrt {{ k\over k +1 }}:$
\eqn\eigenstates{\eqalign{
 |k,m, j_+>    &=
{1\over \sqrt{2k+1}}
( {\sqrt{k+1}} |k,m,0>   + {\sqrt k} |k,m,1> ) \cr
 |k,m, j_-> & =
{1\over \sqrt{2k+1}}
(  -  {\sqrt k} |k,m, 0> + {\sqrt {k+1}} |k,m, 1>) . \cr }
}
The $R^2$ eigenstates when expressed in terms of  spherical harmonics take the
form:
\eqn\rdef{\eqalign{
 |k,m, +>
= &{1\over \sqrt{(2k+2)(2k+3)}}
\biggl\lbrack \sqrt{(k+m+1)(k+m+2)}  Y_{k+1}^{m+1}|1,-1> \cr
- \sqrt{2(k+m+1)(k -m + 1)}
& Y_{k+1}^{m}|1, 0>
 + \sqrt{(k- m+1)(k - m+2)}  Y_{k+1}^{m- 1}|1,1> \biggr\rbrack
\cr
|k,m , ->
= & {1\over \sqrt{2k(2k-1)}}
\biggl\lbrack \sqrt{(k- m)(k-m-1)}  Y_{k-1}^{m+1}|1,-1> \cr
+  \sqrt{2(k+m)(k -m)}
& Y_{k- 1}^{m}|1, 0>
+ \sqrt{(k+ m - 1)(k + m)}  Y_{k-1}^{m- 1}|1,1>  \biggr\rbrack
\cr
|k,m,0>  = & Y_k^m |0,0> \cr
|k,m,1>   =   & {1\over \sqrt{2k(k+1)}}
\biggl\lbrack \sqrt{(k- m)(k+ m+1) } Y_{k}^{m+1}|1,-1> \cr
+ & m {\sqrt 2} Y_{k}^{m}|1, 0>
- \sqrt{(k+ m)(k - m+1 )}  Y_{k}^{m- 1}|1,1>  \biggr\rbrack
\cr
}}
where $|1,1>$, $|1,0>$, $|1,-1>$ are the
spin-one triplet of $R^2$.

We now proceed to express the various  operators appearing in the Dirac
equation {\diracb}
as matrices in the basis
$\big ( |k, m, + >$, $|k, m, j_+>,$ $|k, m, j_->,$ $|k, m, -> \big )$ in which
$K^2$, $K_z$, $L^2$, and $J^2$ are diagonalized.
Since $\sdn$  commutes with $J^2$ and squares to unity,
 after  appropriate normalization of   states, one finds that  $\sdn |k,m, \pm
> =  |k,m, j_{\pm}>$.
 Using this result, one can show that the transverse  Dirac operator   $\DT =
 (J^2 - L^2 - S^2 +1 ) \sdn $ takes the form
\eqn\spheredirac{\DT =
\pmatrix{
0 & - k-1 & 0 & 0 \cr
  k+1 &0&0&  0 \cr
  0 &  0 &  0 &  -k  \cr
0&  0  &  k &  0\cr}. }
 Finally, one can
 determine the matrix form of   the operator  $\nst$ appearing in the gauge
field term in the Dirac equation {\diracb}. Since it  is antisymmetric in
$\vec\s$ and $\vec\t$, it  must  interchange the $r=0,1$ states, and because of
the factor  $\vec n$,  it    changes the orbital angular momentum by one.
Further calculation shows
\eqn\nstmatrix{
\nst = {2i\over 2k+1} \pmatrix{
0 & k +1 &  - {\sqrt{ k(k+1)}} &0 \cr
 - k -1 & 0& 0&{\sqrt {k(k+1)}} \cr
{\sqrt {k(k+1)}} & 0  &0 & - k \cr
0  &  - {\sqrt {k(k+1)}} &  k  &0 \cr
}.}
Substituting {\spheredirac} and {\nstmatrix} in {\diracb}, the Dirac equation
reduces to four coupled linear first-order equations.
It appears that  higher mode zero energy  bound states are forbidden since the
wave function diverges at $r=0$.  Near the horizon,   $r=r_H$, $A=\a
(r-r_H)^{1/2} (1 + O(r-r_H))$ and $B = \b (r-r_H)^{-1/2}(1 + O(r-r_H)) $
while $K$ is constant. Therefore, to leading order in $r-r_H$,
the Dirac equation becomes
\eqn\horizondirac{
\dot \psi + {\a\over \b}   \sdn  \big\lbrack   (r-r_H)   \psi^{\pr}  + {1\over
4} \psi \big\rbrack = 0
}
with solutions $\psi \sim e^{-iEt} (r-r_h)^{({\pm i E\beta /\alpha} - 1/4)}$
which as before diverge at the horizon.

{\newsec{Higgs Field}}
In  this section, we consider the effect of a Higgs field. The Dirac equation
then becomes
\eqn\diraceqh{
i\g^{\m} (\nabla_{\m} -i g A_{\m}) \P - \Phi \P =0
}
where $\Phi = \phi^i \t^i$ is the Higgs field in the adjoint representation. We
have  absorbed the  Yukawa
coupling constant in $\Phi$.
Consider the following spherically symmetric ansatz for $\Phi$
 \eqn\higgs{
\phi^i = F(r) n^i.
}
Since the two chiralities of  the fermion, $\psi_L$ and $\psi_R$, no longer
decouple,   the Dirac equation  now
reduces to  four coupled first order equations  in the $s$-wave sector and  to
eight  in the higher mode sector. Using the fact that $\tdn \chi_1 = -\chi_2$
and    $\tdn \chi_2 = -\chi_1$, we find that the $s$-wave equations become
\eqn\hswaveequations{\eqalign{
{\p f_{R(L)}\over \p t} & \pm  \bigg\lbrack {\p g_{R(L)}\over \p r*} +
 {AK\over r} g_{R(L)} \bigg\rbrack  - i A F g_{L(R)} = 0\cr
{\p g_{R(L)}\over \p t} & \pm  \bigg\lbrack {\p f_{R(L)}\over \p r*} - {AK\over
r}     f_{R(L)} \bigg\rbrack
  -  i A  F f_{L(R)}  = 0 \cr
}}
where $\P_{R(L)} = r^{-1} A^{-1/2} f_{R(L)} \chi_1 +  r^{-1} A^{-1/2} g_{R(L)}
\chi_2 .$
There is in fact still a zero-energy bound state solution to these equations:
\eqn\hzeromode{
f_R  = - i f_L=  \exp{  \int_{r_0}^{r} B({K \over r} - F) \, dr},\quad
g_R=g_L=0
 .}
  Moreover, since the Higgs field causes the wave-function to decay
exponentially at infinity, $\P \sim e^ {- F(\infty) r}$,   there is a bound
state for both odd and even $n$.
As before, for a spacetime with horizon the zero mode diverges there. The
$s$-wave equations can be written as decoupled   second order equations
 \eqn\hsecondordereqn{
{\p^2 f_R\over \p t^2} -{\p^2 f_R\over \p r^{*2}}
+  V(r) f_R =0, \quad f_L= if_R ,\quad  \dot g_R = i \dot g_L = - {\p f_R \over
\p r*} + H f_R }
where
\eqn\heffpotl{
V(r) = {\p H\over \p r^*} + H^2, \quad H \equiv A({K\over r} -F)
.}
 The equations for the higher modes can be found as well.
 The matrix associated with the operator  $\tdn$ appearing in the Higgs field
term is  obtained from $\sdn$ by exchanging $\s$ and $\t$
and using the fact that the $r= 0, 1 $ states are antisymmetric and symmetric
in $\s$ and $\t$ respectively:
\eqn\tdnmatrix{ \tdn =
{\pmatrix { 0 & - {\rm cos}\, 2\xi & \sin 2\xi & 0 \cr
 - {\rm cos}\, 2\xi & 0 & 0 &  \sin 2\xi \cr
 \sin 2 \xi  & 0 & 0 &  {\rm cos}\, 2\xi\cr
0 &  \sin 2\xi  &   {\rm cos}\, 2\xi & 0 \cr
}}
}
 where  $ {\tan}\, \xi = \sqrt {{ k\over k +1 }}.$

{\newsec{Fermion No-Hair Theorem}}
Assuming that the endpoint of gravitational collapse is a stationary black
hole, then the uniqueness theorems imply it must be Kerr-Newman.
  In general, the no-hair theorems  assert
that the only external fields that a black hole can generate
are those yielding  conserved charges in the form of  a surface integral at
infinity. If one attempts
to find a neighbouring solution with hair, the solution for
the perturbation will necessarily diverge
at the horizon of the black hole (assuming it vanishes
at spatial infinity.)
 We should point out that a Yang-Mills black hole such as the one discussed
earlier does not unambiguously violate the no-hair theorems since the
non-linear nature of the  field allows one to view  it  and not the black hole
as being the source of the external Yang-Mills field.
{}~\footnote{*}{\rm In fact, Einstein-Skyrme black holes with Yang-Mills hair
have been constructed recently{\ref\droz{S. Droz, M. Heusler, and N. Straumann,
{\it Phys. Lett. } {\bf B268} (1991) 371.
}}, and are stable under linear perturbations{\ref\droz {   M. Heusler, S.
Droz, and N. Straumann, {\it Phys. Lett. } {\bf B285} (1992) 21.}}.}
After all, one would certainly ascribe the complicated gravitational field of
an accretion disk surrounding a black hole to the disk and not to the black
hole.  The higher multipoles moments of the gravitational field can be
unambiguously identified with the higher moments of the matter density
distribution of the disk. The Yang-Mills field, however, does not have well
defined multipole moments. In particular,
attempts to construct a total charge by integrating the Yang-Mills magnetic
field does not yield a gauge invariant quantity since there is a free
uncontracted  group index.

The fact that the zero mode {\zeromode}   ({\hzeromode}) diverges on the
horizon is   expected from these no-hair theorems. In this section, we prove a
general no-hair theorem for fermions. This has already been done for
Schwarschild in
{\ref\hartle{J. Hartle, in  {\it Magic without Magic}, ed. J. Klauder (San
Francisco: Freeman,  1972).}}
 and \ref\teit{C. Teitleboim, {\it Phys. Rev. D} {\bf 5} (1972) 2941.}
 By employing the conformal invariance
of the Dirac equation, we show that all static  solutions in a spherically
symmetric spacetime diverge at the horizon.
 Any static, spherically symmetric spacetime may be written in isotropic
coordinates
\eqn\isometric{
ds^2 = - V^2 dt^2 + W^2 d{\bf x} \cdot d {\bf x}
}
where $V$ and $W$ are functions of $|{\bf x}|$. (This ansatz
is, in fact, more general than being spherically symmetric
applying to multi-black hole metrics as well.)
Consider the gauged Dirac equation
\eqn\gaugeddirac{
i\g^{\m} ( \nabla_{\m} - i g A_{\m} ) \Psi - m \Psi = 0
}
where $m$ might depend on a Higgs field and hence on position.
In $d$ spacetime dimensions we have the following  result:
if $(\Psi, g_{\m\n}, A_{\m}, m)$ is a solution of {\gaugeddirac},
then  $(\O^{d-1\over 2} \Psi, \O^{-2} g_{\m\n}, A_{\m}, \O m )$
is also a solution. For  the metric
{\isometric} with  $A_0=0$, this implies
that $V^{3/2} \P$ solves {\gaugeddirac} with mass term
$Vm$ in the ultrastatic optical metric
\eqn\ultrastatic{
ds^2 = - dt^2 + {W^2\over V^2} d{\bf x} \cdot d {\bf x}
.}
As is the case for no-hair theorems,  we are interested in time-independent
solutions
 to the Dirac equation.
 Applying
conformal invariance again, but now to the spatial part of the metric implies
that
$ \chi\equiv V^{1/2} W \P$ solves the flat three-dimensional gauged
Dirac equation with mass term $m W$.

 For $m=A_i = 0$ the flat space Dirac equation may
be solved in a variety of ways, but
perhaps the most illuminating from the present conformal viewpoint is
to note that in flat Euclidean $n$-space,
\eqn\harmonic{
\chi^{\a} = T^{\a}_{i_1 i_2\ldots i_l} x^{i_1}x^{i_2} \ldots x^{i_l}
}
clearly solves the flat space Dirac equation with $m=A_i =0$
provided
$ T^{\a}_{ij\ldots k}$ satisfies
\eqn\tsoln{
\g^{i\a}_{\;\; \b } T^{\b}_{ij\ldots k} = 0
.}
Solutions which vanish at spatial infinity can be obtained
from {\harmonic} by an inversion.
To each  static solution $\chi (x^i)$ to the flat
space Dirac equation in $n$ spatial dimensions,
there is an  inverted solution
${\g_i x^i \over r^n} \chi (x^i/r^2).$
Applying this to {\harmonic}, we obtain the general
multipole solution
\eqn\flatmultipole{
\chi = {\g_i n^i \over r^{l+2}}
 T_{i_1 i_2\ldots i_l} n^{i_1}n^{i_2} \ldots n^{i_l}
}
where $n^i = x^i/r$.
Thus, if $m=A_i=0$, the general solution in the spacetime
 {\isometric}
which decays at infinity is given by
\eqn\multipole{
\Psi = V^{-1/2} W^{-1} {\g_i n^i
\over r^{l+2}} T_{i_1 \ldots i_l} n^{i_1} \ldots n^{i_l}
.}
It is now clear that $\P$ will blow up at a non-extreme horizon
for which $V=0$ for $r>0$. In the extreme case for a regular horizon
where
$W \ra r^{-1}  $ and $ V \ra r$,
we see that $\P$ still blows up for any permitted value of $l$.

\bigbreak\bigskip\bigskip\centerline{\bf Acknowledgements}\nobreak
We thank N. Manton  for many useful discussions.

  A.S. would like to acknowledge   the support of the SERC.

\listrefs
\end